\documentstyle[eqsecnum,floats,preprint,aps,psfig]{revtex}

\tighten

\begin{document}

\setlength{\oddsidemargin}{0.0truein}
\setlength{\textwidth}{6.0in}
\setlength{\textheight}{8.5in}
\setlength{\topmargin}{-0.50in}

\def\a{\alpha}
\def\b{\beta}
\def\g{\gamma}
\def\d{\delta}
\def\e{\epsilon}
\def\h{\eta}
\def\th{\theta}
\def\bth{\bar{\theta}}
\def\y{\psi}
\def\by{\bar{\psi}}
\def\p{\pi}
\def\o{\omega}
\def\f{\phi}
\def\F{\phi}
\def\x{\xi}
\def\m{\mu}
\def\s{\sigma}

\def\ds{\sum_{i}}
\def\la{\langle}
\def\vp{\varphi}
\def\baF{{\tilde{F}}_{\beta}}
\def\fd{\phi ^{\dagger}}
\def\t{\tilde} 
\def\lm{\lambda}

\newcommand\bg{\begin{eqnarray}}
\newcommand\ed{\end{eqnarray}}
\newcommand\bgn{\begin{eqnarray*}}
\newcommand\edn{\end{eqnarray*}}

\def\D{\partial}
\def\del{\partial}
\def\ra{\rightarrow}
\def\phat{\hat{p}}
\def\qhat{\hat{q}}
\def\Lhat{\hat{L}}

\def\ms{m^{*}}
\def\mstwo{ {{m^{*}}^{2} }}

\def\Norm{{\cal{N}}}

\def\ephi{ \left< \> \phi \> \right> }
\def\muhat{\hat{\mu}}
\def\pfour{\phi^{4}}
\def\pren{{\phi}_{R}}

\def\inf{\infty}
\def \Z{ Z[ J] }
\def \Zu{Z^{(ul)}}
\def \Su{S^{(ul)}}
\def \vacl{< \> 0 \>\vert \> }
\def \vacr{\> \vert \>0 \> >}
\def \derij{ {\delta \over \delta J(x )}}

\def \dih{{\,\cal{D}}_{N}\,}
\def \cyc{{\,\cal{C}}_{N}\,}


\input epsf


\pagenumbering{arabic}

\begin{titlepage}

\title{\bf{Theta Vacua and Boundary Conditions of the Schwinger-Dyson 
Equations}}


\author{
Santiago Garc\' \i a$^{1}$\footnote[1]{sgarcia@watson.ibm.com},
Zachary Guralnik$^{2}$\footnote[2]{zack@puhep1.princeton.edu},
and G.S.~Guralnik$^{3}$\footnote[3]{gerry@het.brown.edu.~~Permanent Address: 
Physics Department, Brown University, Providence RI. 02912.}}

\address{~\\$^1$T.J. Watson Research Center, Yorktown Heights, NY 10598}

\address{~\\$^2$Joseph Henry Laboratories, Princeton University, 
	        Princeton, NJ 08544}

\address{~\\$^3$Center for Theoretical Physics, Laboratory 
		for Nuclear Science\\
		Massachusetts Institute of Technology,
                Cambridge, Massachusetts 02139}

\maketitle


\begin{abstract}
\noindent

Quantum field theories and Matrix models have
a far richer solution set than is normally considered,  
due to the many boundary conditions which must be set to specify
a solution of the Schwinger-Dyson equations.
The complete set of solutions of these equations is obtained by 
generalizing the path integral to include sums over various inequivalent 
contours of integration in the complex plane.  
We discuss the importance  of these exotic solutions. 
While naively the complex
contours seem perverse,  they are relevant to the study of theta vacua
and critical phenomena.  Furthermore, it can be shown that within 
certain phases of many theories,  the physical vacuum does not 
correspond to an integration over a real contour. 
We discuss the solution set for the special case of 
one component zero dimensional scalar field theories,  and
make  remarks about matrix models and higher dimensional field theories
that will be developed in more detail elsewhere. 
Even the zero dimensional examples have much  structure,  and
show some analogues of phenomena which are usually 
attributed to the effects of taking a thermodynamic limit.

\end{abstract}

\thispagestyle{empty}

\vskip-20cm
\noindent
\phantom{bla}
\rightline{}
\rightline{MIT-CTP-2582}
\rightline{PUPT-1670} 
\end{titlepage}

\section{Introduction}
This paper is a consequence of our ongoing study of a new numerical
approach to quantum field theory,  known as  
the Source Galerkin \cite{firstpaper,john2,john1} technique.
This method is based on systematically approximating
the functional differential equations,  or Schwinger-Dyson equations, 
satisfied
by the generating functional in quantum field theory,  while
controlling the error with a weighted
averaging procedure.
While exploring this method, it came to our attention
that it is necessary to carefully analyze the boundary conditions imposed
on these equations.

Here, we address this problem from a formal rather than a numerical 
point of view. 
For any theory with a large but finite number of degrees of 
freedom,
there are a very large number of ``theta'' parameters characterizing solutions 
of the Euclidean Schwinger-Dyson equations,  which are the quantum mechanical 
version of the 
equations of motion.   Only one point in this parameter space
corresponds to the usual path integral \footnote{Sometimes no point 
corresponds
to the usual path integral.  This  is the case if the action is unbounded 
below.
Such a situation occurs in Euclidean Einstein gravity and in certain matrix
models}.   It is therefore tempting to discard all the other solutions as 
unphysical.   However, we shall argue that under many conditions the exotic
solutions are physical.   Symmetry breaking vacua 
provide the simplest examples of a situation in which  exotic solutions are
physical.  Furthermore,  critical phenomena have a natural interpretation in 
terms of the behavior of the full solution set in the thermodynamic limit.
In this limit the solutions associated with many different boundary conditions 
coalesce. For other  boundary conditions,  the limit does not exist.  
Which boundary conditions have a thermodynamic limit and which ones coalesce 
depends on the theory's parameters in a manner which determines the phase 
diagram and the critical exponents. 

In section II we discuss the boundary conditions of some simple 
zero dimensional scalar theories.  
In addition to the usual integral solution
of the Schwinger-Dyson equations,   there are exotic solutions which
are obtained from
sums over various inequivalent complex contours of integration. 
Zero dimensional theories have many phases,  all of which are continuously 
related.  
For symmetric actions these include symmetry breaking solutions.

In section III,  we describe the boundary conditions of the Schwinger-Dyson 
equations
for a general theory with a finite number of degrees of freedom.  
For a Euclidean field theory on a finite lattice without a boundary,
the complete set of 
solutions may be obtained from different complex integral representations  
and have 
a simple correspondence to the classical solutions of the theory.
The exotic solutions may be thought of as generalized lattice theta vacua. 
There is a still larger class of solutions which may be obtained on
a lattice with a boundary,  at which the Schwinger-Dyson equations may 
be modified.  In the integral representation of the solution,  the measure 
for fields at the boundary is essentially arbitrary,  and may be set
by surface terms in the action or by wave functions.  However  we will confine 
our discussion to generating functionals in the vacuum sector.  These 
may be obtained on a periodic lattice, for which the exotic contours
provide a complete set of boundary conditions.
In the thermodynamic limit the real path integration with surface terms in
the action  gives some but not all of the solutions 
found using complex contours for a periodic lattice\footnote{Here we
assume that for real fields the potential is not unbounded below.}.
There are compelling reasons to 
consider solutions represented by exotic contours, of which we list a 
few below.   One reason is that certain aspects of critical 
phenomena are very naturally 
related to the behavior of the full set of solutions, including those 
which are unphysical, in the  thermodynamic limit.  The full solution 
set may only by
obtained by allowing complex contours.
Furthermore solutions corresponding to false vacua may be  
of physical interest but are not obtainable from the integration
over real fields.  In the case of matrix models it is essential to consider
exotic contours to obtain even certain physical solutions with a real double 
scaling limit.  This is true even when the matrix potential is bounded below
for real eigenvalues.
In this case there is no analogue 
of a spatial boundary at which the measure may be arbitrary. 
There are often many physical 
solutions having a real double scaling limit.  One can obtain some 
of these  solutions from the integration over real eigenvalues  
by making global 
perturbations of the action which are removed after taking the 
$N\rightarrow \infty$ or double scaling limit.  However there
are also examples of solutions
which can not be found this 
way\cite{CI}\cite{zackparis}\cite{nextmmpaper}.         
On the other hand,  the complete set of exotic contours does yield 
the full solution set\cite{zackparis}\cite{nextmmpaper}\cite{davide}.  
There are 
situations in which the boundary conditions associated with a 
real solution in the double scaling limit are extremely 
unusual\cite{zackparis}\cite{nextmmpaper}.

In section IV, we discuss phenomena which occur in the thermodynamic limit in 
the context of the Euclidean Schwinger-Dyson equations.  
In particular, we discuss the 
coalescence of solutions associated with different boundary conditions,  and
the absence of a thermodynamic limit for other boundary conditions.  
In some cases
one can explicitly see how the solution set collapses,  
as in certain large $N$ matrix models.  
The solutions which survive in this limit do not necessarily coalesce 
with a solution associated with the integral over real fields,  and sometimes 
the integral over real fields itself does not have a thermodynamic limit. 
In this paper we will not explicitly 
demonstrate the collapse of the solution set.  Instead we will assume the
collapse and show that it leads to several expected phenomena,  such as
the appearance of phase boundaries in the thermodynamic limit due to the 
accumulation
of Lee-Yang zeroes.  Critical exponents are determined by the manner in 
which the solution set collapses.  
We give a simple zero dimensional analogy in 
which tuning a coupling
constant to zero shrinks the solution set and causes Lee-Yang zeroes to 
accumulate. 
Setting the coupling constant to zero in this example is analogous to 
taking the 
thermodynamic limit.  We also show that the collapse of the solution set 
induces another set of well known equations which we will refer to as the
Schwinger action principle\footnote{The Schwinger action principle can 
be used to produce all 
equations associated
with the formulation of a quantum field theory including the 
Schwinger-Dyson equations. 
However, as a matter of convenience, we will refer to equations exclusive 
of the
Schwinger-Dyson equations as the Schwinger action principle or 
just the 
action principle.};  
\bg
{\del\over\del g} Z = <{\del S \over \del g}> Z
\label{SDY}
\ed
Here $g$ is a parameter of the the action $S$,  and $Z$ is the partition 
function.  
If one only considers the usual solution of the Schwinger Dyson equation 
corresponding to the integral over real eigenvalues in a matrix model,  or
real fields in a field theory on a periodic lattice,
then the action principle is obviously obeyed. 
However, as soon as one considers all of the
solutions of the Schwinger-Dyson equations, it is no longer obvious that
the action principle should be satisfied.
For a 
finite system the action principle is independent of the Schwinger-Dyson 
equations. It is possible to satisfy the Schwinger-Dyson equations  while 
violating the action principle.
Since the solution set does not necessarily collapse to the usual integral 
solution, 
it is nontrivial to show that the action principle follows from the 
Schwinger-Dyson
equations in the thermodynamic limit.   

In section V we discuss the effect of the collapse of the 
solution set on the form of the effective potential.  For a finite number of 
degrees of
freedom,  there are an infinity of effective potentials,  each of which has 
only 
single extremum.  In the thermodynamic limit,  it is possible to have a single
effective potential with multiple extrema.

\section{Boundary conditions of Schwinger-Dyson equations}

When there is no spatial boundary, 
the set of boundary conditions of the Schwinger-Dyson equations
is determined by the behavior of the action at asymptotic values
of the fields.  The reasons for this will become clear shortly.
For many theories a potential term dominates the action
at large fields,  and we will begin by considering a theory with only a 
potential term.
Let us find the space of solutions for a polynomial
action with one degree of freedom,  
\bg
S= \sum_{l=1}^n {1\over l} g_l \phi^l
\label{phi}
\ed 
The generating function $Z(J)$ of disconnected Green's functions satisfies 
the Schwinger-Dyson equation:
\bg
\sum_{l=1}^{n}(g_l{\del^{l-1}\over \del J^{l-1}} - J)Z(J) = 0
\label{sdyson}
\ed 
This is a linear differential equation of order $n-1$,
and has an $n-1$ parameter set
of solutions.  One parameter corresponds to a trivial normalization factor,  
so  
there is an $n-2$ parameter set of solutions with different
Green's functions.
There is a complete basis set of solutions with the 
integral
representation 
\bg
Z(J) = \int_{\Gamma} \,d \phi  {\displaystyle{e^{-S(\phi) + J \phi}}}
\label{complexcont}
\ed
where $\Gamma$ is a complex path such that
\bg
\int_{\Gamma} \,d \phi  {\del \over \del \phi}{\displaystyle{e^{-S(\phi) + 
J \phi}}} = 0
\label{bterm}
\ed
or
\bg
\left.{\displaystyle{e^{-S(\phi) + J\phi}}} \right|_{\partial \Gamma} =0 .
\label{boundterm}
\ed
The paths for which this is true is determined by the order of the 
polynomial potential.
We require $Re (g_n \phi^n) \rightarrow +\infty$ 
as $| \phi | \rightarrow \infty$, giving $n$ domains in 
which the contour can run off to infinity (fig.1).   One can then write 
down $n-1$ independent contours
which comprise a basis set.  
The contours may not be deformed into one another 
since $e^{-S}$ is not analytic at $| \phi | \rightarrow \infty$. 
For $S= {1\over 4}g\phi^4+...$ with $g$ real,  one choice of independent 
contours is 
\bg
\Gamma^+ = [-\infty, 0] + [0, +i\infty]\cr
\cr
\Gamma^0 =  [-\infty, 0] + [0, +\infty]\cr 
\cr
\Gamma^- = [-\infty, 0] + [0, -i\infty]    
\label{gammas}
\ed
with which we associate the  generating functions $Z^+, Z^0$ and $Z^-$.
As a matter of definition, we will say that solutions at different values of 
$g$ 
are in the same phase if they are related by the the Schwinger action 
principle,
\bg
( {\del\over\del g} + {1\over 4} {\del^4 \over \del J^4} ) Z = 0.
\label{wha}
\ed
For the example above,
a general solution of the Schwinger-Dyson equations is of the form
$Z = aZ^+ + bZ^0 + cZ^-$  where $a$, $b$ and $c$  may be  
arbitrary functions of $g$.  
The solutions $Z^+$, $Z^0$ and $Z^-$ 
separately satisfy the action principle.  Therefore the action principle 
is satisfied
if $a$,$b$ and $c$ are independent of $g$. A phase is labeled by the 
ratios of these three constants.  However as $Re(g)$ becomes negative, 
the contours $\Gamma^+, \Gamma^0$ and $\Gamma^-$ are no longer convergent. 
For $g$ real and positive,  there are domains swept out by contours 
equivalent to
$\Gamma^+, \Gamma^0$ and $\Gamma^-$. As one rotates the phase of $g$, these 
domains also rotate to maintain convergence.  
So to be more precise the action principle is
satisfied if for infinitesimal variations
of $g$ there is a choice of contours which is held fixed.  If one makes 
large changes in $g$,  it may be necessary to change the contour.  Thus the 
statement
that two solutions are in the same phase if they are related by the action 
principle
makes sense only locally in the space of coupling constants.  
A large loop in coupling constant space can go around
a phase boundary and onto another Riemann sheet,  thus changing the boundary
condition.  In the above example there is a phase boundary at $g=0$, and a 
$2\pi$ rotation of $g$ forces a ${\pi\over 2}$ rotation of the contour to 
maintain 
convergence.  Note that in zero dimensions the Schwinger action principle 
is a very
arbitrary way to define a phase.   One could have required the generating
function within a given phase to satisfy  any other set of first order 
differential 
equations in the couplings.  As we shall see later,  the action 
principle is a consequence of the Schwinger-Dyson equations in the 
thermodynamic limit \footnote{Even in zero dimensions,  if one solves the
Schwinger-Dyson equations in a weak coupling expansion,   the action 
principle is 
automatically satisfied order by order.  By performing a weak
coupling expansion,  one has chosen a restricted 
class of phases with a particular weak coupling asymptotics.}.  

\begin{figure}

\centerline{
\epsfxsize=9cm
\epsfbox{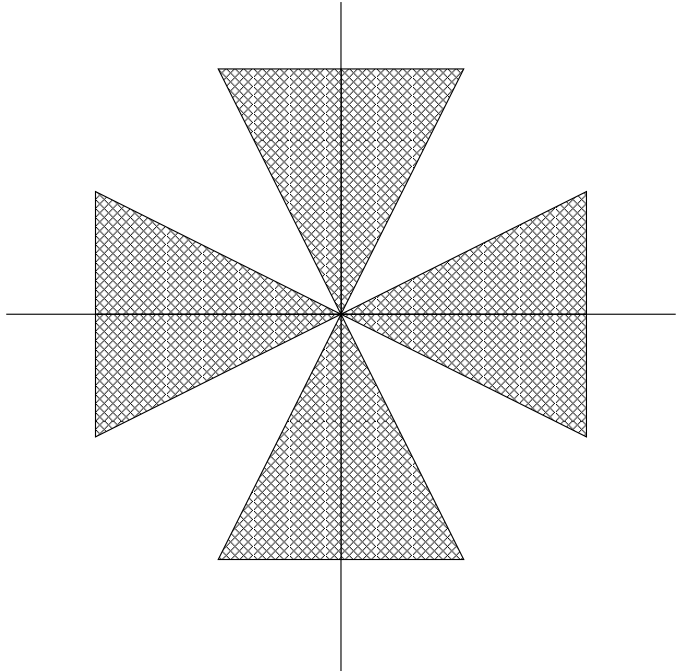}  }

\caption{The shaded area 
is the region of the complex $\phi$ plane defined by  
$\cos(4 \theta) > 0$, where
$\theta$ is the argument of $\phi$. Any path starting and ending at 
infinity within one of these four domains 
corresponds to a particular
solution for the zero dimensional $\phi^4$ theory}
\label{fig1}
\end{figure}

For the example of the action $S= {g\over4}\phi^4 + {1\over 2} \mu\phi^2$  
the exact 
disconnected Green's functions (see Appendix I) satisfying both 
the Schwinger-Dyson equations and the action principle are given by;
\bg
G_{2n} = \mu^{-n}(2n - 1)!! t^n 
	{  U(n,t) + (-1)^{n} \rho U(n, -t) \over
	   U(0,t) + \rho U(n, -t)     } 
\label{c}
\ed
and 
\bg
G_{2n+1} = \mu^{-n-{1 \over 2} }{2n !!\over n!} (-t)^n 
		{V(n+{1 \over 2},t)\over V( {1 \over 2} ,t  )  }
	   \alpha {  t^{1\over 2} e^{t^2\over 4} \over
	        U(0,t) + \rho U(0,-t)  } 
\label{d}
\ed
where $t=\sqrt {\mu^2 \over 2 g}$,  and $U(a,t)$ and $V(a,t)$ are 
parabolic cylinder
functions defined in appendix I.
The parameters  $\alpha$ and $\rho$ depend on the ratios of the 
coefficients, $a$, 
$b$ and $c$,  of the
independent contours.
By requiring Green's functions to be real,  or that $a$ be real and $b = c^*$,
one is still left with many exotic solutions. 
These are labeled by the two real parameters $\alpha$ and $\rho$.
Solutions with $\alpha \ne 0$  break the $\phi\rightarrow -\phi$ symmetry of 
the action.
Such solutions exist because the contours $\Gamma^+$ and
$\Gamma^-$ break the symmetry. 

Since reality is not much of a constraint on the solution set, it 
might be tempting to require positivity of even
Green's functions as well.  This is a much stronger constraint,  since only 
the 
solution corresponding to the real contour is manifestly positive.  For a 
general boundary
condition,  one must do some work to ascertain positivity.  However, it does 
not 
make sense to impose positivity before taking a thermodynamic limit,  
because the conditions
for positivity may change in this limit.  Our approach to the 
selection of boundary conditions
will be to
let the thermodynamic limit do most of the selection
for us.  We shall argue later that the solution set collapses in the 
thermodynamic limit.
By this we mean that the solutions associated with certain continuous 
classes of  boundary 
conditions coalesce,  while other boundary conditions lead to no 
thermodynamic limit. 
The surviving solutions may then include some which violate 
reflection positivity
(for Euclidean Schwinger-Dyson equations),  and  these may be 
discarded by hand.  
Note that sometimes complex solutions describe
false vacua and are of physical interest.  
	
One might also try to constrain the solution set by requiring that the
weak coupling limit smoothly approach the free field solution.  
This would mean that $Z$ should have the formal
representation
\bg
Z(g,J)= e^{ {g\over 4} \del_J^4}Z(0,J)
\label{weak}
\ed
This is a very strong constraint reflecting the fact that the Schwinger-Dyson
equation becomes first order at $g=0$.  For the zero dimensional $\phi^4$ 
theory the equation is   
$(\mu\del_J - J)Z = 0$ and, consequently,
has only one solution up to a normalization.  
However, this constraint is ill defined if there are any phase boundaries
in $g$.  In our zero dimensional example there is a branch point at $g=0$,  
so the result of taking a weak coupling
limit depends on the path.  One could take $g\rightarrow 0$ with the 
phase of $g$ fixed.
There is no physical reason why the partition function should 
have a free field 
limit as $g\rightarrow 0$.  In the symmetry breaking phase of a $g\phi^4$ 
theory, 
the Green's functions are singular at $g=0$.  For instance, the one point 
function at 
weak coupling is $<\phi> = \sqrt{ -{\mu\over g} }$.
For our zero dimensional example,  
the requirement of 
a free field $g\rightarrow 0$ limit with $\mu<0$  would yield the symmetric 
boundary condition 
corresponding to an integral over the imaginary axis.   
 
It is enlightening to consider the weak coupling limit of the full set of 
solutions of 
the zero dimensional
$\phi^4$ theory.  There are some solutions to 
which the effective potential in the loop expansion
is asymptotic.  These correspond to sums over contours 
which pass through a single dominant saddle point.
At each order in the loop expansion the effective 
potential has three extrema corresponding to 
each saddle point  and a different class of boundary 
conditions \footnote{The three extrema are a spurious feature of the 
loop expansion in
zero dimension.  In fact non-perturbatively there is a different 
effective potential 
for every boundary condition. This will be discussed in a later section.}.
Among these are symmetry breaking solutions with the asymptotic 
expansion for the one point 
function given by
\bg
<\phi>= \sqrt{-{\mu\over g}}(1+O(g) +\dots)
\label{wk}
\ed
$Z^+$ is one example of a solution with this expansion.  Another 
example which is real is 
$Z^+ + Z^-$ for $g>0$ and $\mu<0$. 
One can explicitly show that such a boundary 
condition leads to the symmetry breaking phase of
a $Tr\phi^4$ matrix model\cite{zackparis}\cite{nextmmpaper}\cite{davide}.
It seems natural to conjecture that some generalization of
such a boundary condition also corresponds to the symmetry breaking 
phase of a $\phi^4$ field theory.

There is yet another class of 
solutions of the zero dimensional $\phi^4$ theory 
which does not correspond to the loop expansion.
These  arise when the contours cross degenerate dominant
saddle points.  These solutions are linear combinations of the ones 
associated with 
the loop expansion.  Some of these are 
extremely singular at weak coupling.  For instance for 
$\mu>0$ and $g>0$,  the solution $Z= Z^{+} + Z^{-}$  has a one point function
with the expansion
\bg
<\phi> = \sqrt{ {-\mu\over g}} e^{+{\mu^2\over 4g} }(1+\dots)
\label{sing}
\ed
This is also the expansion for a similar solution in the case $\mu <0$,  
but the boundary condition is different,  with contours rotated 
by $\pi\over 2$.  
The essential singularity $e^{+\mu^2\over 4g}$ arises
because $<\phi> = {\del_J Z\over Z}$,  and the contribution of the 
two degenerate dominant
saddle points at $\phi = \sqrt{{-\mu\over g}}$ cancels in the denominator 
but not in the 
numerator.  In one matrix models it is easy to see that boundary 
conditions of this
singular type do not have a thermodynamic limit \cite{zackparis}\cite{nextmmpaper}.  
However at this point of our discussion we already have some evidence 
that some the other exotic boundary 
certainly may be physical,  the simplest example being
that of the symmetry breaking solutions associated with the loop expansion.

Our interest in $\phi^4$ theory is primarily as a toy model.
The same analysis may be applied to obtain the solution set for 
non-polynomial actions,  in which  the order of the Schwinger-Dyson
equations is not always  obvious.
A simple example is given by the loop equation of one plaquette QED,
which is defined by the action 
$S = -\beta cos \phi$.  The condition 
that $e^{-S}$ vanishes as $|\phi| \rightarrow \infty$ restricts 
the domain of integration to satisfy $Re (\beta cos \phi) \rightarrow +\infty$ 
asymptotically.
For $\beta$ real and positive,  one choice of inequivalent contours which 
satisfy this requirement is given by
\bg
\Gamma_1 = [-i\infty, +i\infty]
\label{kal}
\ed
and
\bg
\Gamma_2 =[-i\infty, 0] + [0, 2\pi] + [2\pi, 2\pi + i\infty]
\label{cal}
\ed
to which we associate the partition functions $Z_1$ and $Z_2$.
Due to the periodicity in $Re( \phi)$, it is easy to see that
any other allowed contour may be 
written as linear combination of these two.    
The difference between $Z_1$ and $Z_2$  
is just the usual integral defined on the compact interval $[0,2\pi]$. 
Note that only this solution manifestly satisfies the 
condition that $|<e^{i\phi}>| < 1$.
However along the lines of  our previous arguments, we will not 
impose such conditions
before taking a thermodynamic limit.

Given that there are two independent contours, we expect 
that the Schwinger-Dyson  equations should be second order.
In this case, it is easy to construct the equations even
though the action is not polynomial.  By coupling sources $J$ and
$\bar J$ to the loop variables $e^{i\phi}$ and $e^{-i\phi}$ one obtains the 
Schwinger-Dyson equations
\bg
[\beta(\del_J - \del_{\bar J} ) - 
2(J\del_J - {\bar J}\del_{\bar J} )]Z(J,{\bar J}) = 0
\label{plaq}
\ed
and the constraint
\bg
\del_J  \del_{\bar J}Z(J,{\bar J}) = Z. 
\label{pha}
\ed
These equations may be rewritten in the form
\bg
{\beta - 2J \over \beta - 2{\bar J}} \del_J^2 Z = Z
\label{scn}
\ed
and
\bg
\del_{\bar J} Z = {\beta - 2J \over \beta - 2{\bar J}}\del_J Z
\ed
which,  accounting for the normalization of $Z$,  has a one dimensional 
space of solutions
characterized by the above integral representations.
Of course,  in contrast to a theory with a 
local order parameter, one can not hope to learn very much about the phases
of a gauge theory from a zero dimensional (one plaquette)  model.
Note that lattice gauge theories are examples of
theories in which the large field behavior of action,  and therefore 
the choice of boundary
conditions,  is not controlled by independently variable local terms, 
due to the  
constraints among the plaquettes.
However the one plaquette model does illustrate something which is true 
generally.
The ``exotic''  solutions of the loop equations for a gauge theory
correspond to certain complexifications of the gauge group.  
Similarly, in hermitian matrix models 
the exotic solutions correspond
to certain complexifications of the eigenvalues in the matrix integral, and 
in this case it can be proven
that some of the exotic solutions are ``physical'' for certain values of 
the couplings,
meaning that a real double scaling limit exists.

\section{From Zero to Higher Dimension}

For theories in which independently variable local terms dominate 
the large field behavior of the
action, 
it is straightforward to represent the higher dimensional theory in terms of a
zero dimensional theory.  Such representations are the starting point for both
the strong coupling and mean field expansions.
For example,
consider the lattice Schwinger-Dyson equations of an interacting
scalar field with action
$S= {1\over2}\sum_{x,y}H_{x,y}\phi_x \phi_y + \sum_x V(\phi_x)$:
\bg
(\sum_y H_{x,y}{\del\over\del J_y} + 
 V^{\prime}({\del\over\del J_x}) - J_x)Z[J] = 0 .
\label{latte}
\ed
A formal solution corresponding to the strong coupling expansion is given by 
\bg
Z[J] = e^{ {1\over2} \sum_{x,y}{\del\over\del J_x}H_{x,y}{\del\over\del J_y}} 
	\prod_w Z_0(J_w)
\label{std}
\ed
where $Z_0(J)$ is any solution of the zero dimensional Schwinger-Dyson equation
\bg
(V^{\prime}({\del\over\del J}) - J)Z_0(J) = 0 .
\label{zd}
\ed
The solution also may be written as 
\bg
Z[J] = \int\prod_x d\rho_x 
	e^{-{1\over 2}\sum_{x,y}H^{-1}_{x,y}\rho_x\rho_y}
	\prod_x Z_0(J_x + \rho_x) .	
\label{hub}
\ed
This may be thought of as a stochastically driven zero dimensional theory.
A saddle point expansion in the auxiliary variable $\rho$ yields a 
mean field expansion.  We shall not make any use of these expansions here,  
but just mention them to illustrate the structure of the solution set in 
non-zero dimensions.  In both the representations 
there is a product over 
zero dimensional solutions at each lattice site.  It is not necessary to use 
the same 
integration contour at every lattice site,
even if lattice symmetry is required of the solution.  
A sum over translations and rotations of solutions 
with inhomogeneous boundary conditions yields a solution which satisfies the 
lattice symmetries, and yet can not always be written in terms of an integral 
over 
a product of 
equivalent measures at each lattice site (fig.2).  
In a future publication\cite{nextmmpaper} we show that there are matrix 
model solutions with
this type of boundary condition.  In this case the lattice site is replaced
with an eigenvalue label,  and the integral representation does not involve 
a single  
measure for each eigenvalue\cite{nextmmpaper}.  This means that the 
solution is not 
of the form,  
\bg
Z= \prod_n \left(\sum_i c_i \int_{\Gamma_i}d\Lambda_n \right) 
\Delta^2 [\Lambda]
e^{- N\sum_m V(\Lambda_m)}.
\label{not}
\ed
where $\Lambda_n$ are the eigenvalues, and $\Delta[\Lambda]$ is
the Vandermonde determinant.
If we consider only solutions of this form, which involve products of 
equivalent measures over the
lattice sites,  then the space of boundary conditions on a periodic
lattice is the same
as that of the zero dimensional theory.  

\begin{figure}

\centerline{
\epsfxsize=10cm
\epsfbox{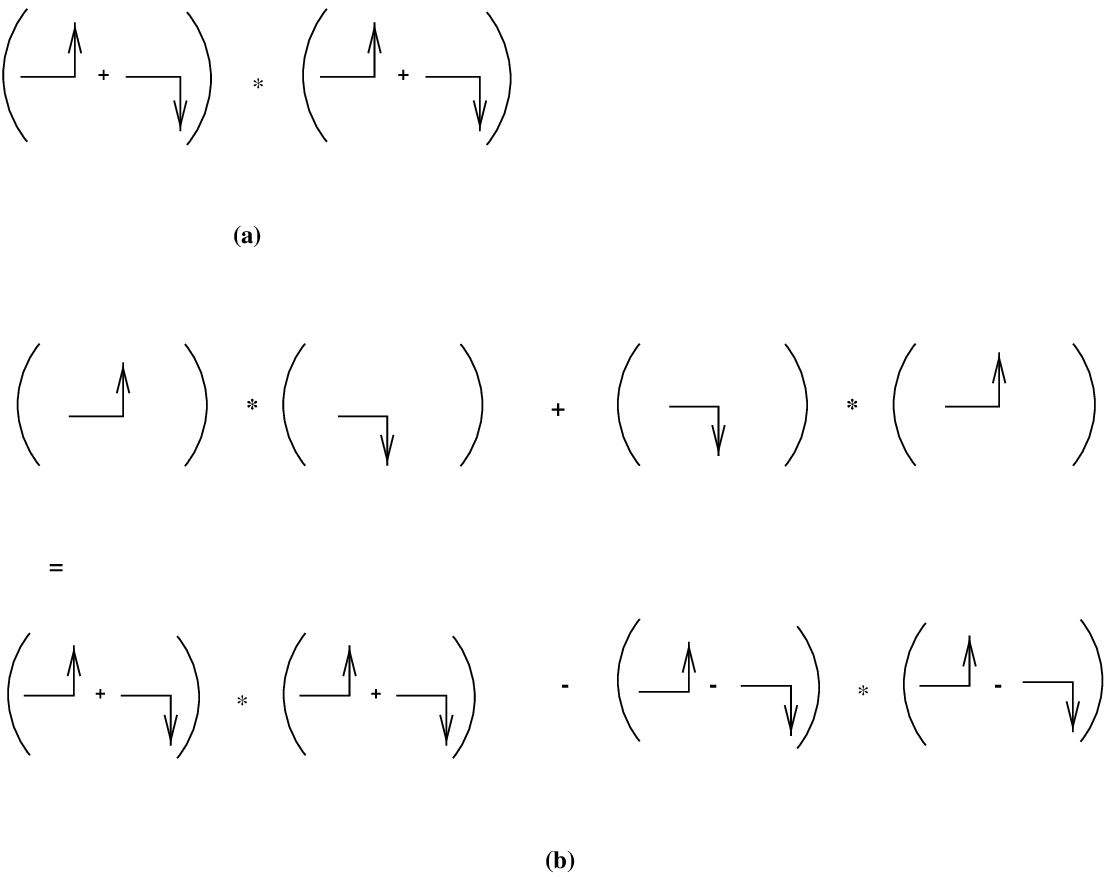}    }

\bigskip
\caption{A pictorial representation of translationally invariant
solutions on a two site lattice.  (a) is a product over equivalent measures
at each lattice site.  (b) may not be written as a product over equivalent 
measures at each site.}
\label{fig2}
\end{figure}

Note, however, that there is a class of theories for which there is 
no simple 
relation between the space of
boundary conditions in a zero dimensional theory and a subspace of 
the boundary conditions on a periodic lattice. 
This class consists of theories in which the action at large values of 
the fields is not
dominated by independently variable local terms. 
We have stated that gauge theories fall in this class. 
Another example of a theory in this class is the XY model,  with 
lattice action
\bg
S= -{1\over 2} H_{x,y}cos(\theta_x - \theta_y).
\label{XYmod}
\ed
For such theories it is usually much harder to de-numerate the boundary 
conditions of
the Schwinger-Dyson equations,  although there is a partial classification 
based on   
generalized lattice 
theta vacua which we discuss shortly.  
It seems likely that there
is an equivalence or at least a large overlap between theories without a 
local order 
parameter,  
and theories for which the boundary conditions are not 
determined by independent local terms. 
Note that for a theory in which local terms control the 
boundary conditions,  such as the
$\phi^4$ theory,  the behavior of the local order parameter $<\phi>$ 
is determined by the
choice
of these boundary conditions.  
If in the thermodynamic limit the solution set collapses,  then certain 
boundary 
conditions and therefore certain values of the order parameter are selected. 
 
To argue that the exotic solutions involving complexified 
fields may in fact be physical,
we have so far given the example of symmetry breaking.  We 
will now try to make our arguments
more concrete.  While the exotic solutions may appear very 
strange,  many  of them
are actually familiar.  
For example, there is a simple relation  between
conventional theta vacua and
exotic solutions with complexified fields,  which we
demonstrate below.  
For some simple cases it can be shown that a complete set of 
solutions of the Schwinger-Dyson equations 
may be generated from the complete set of classical solutions,  
both real and complex. 
This relation may be made explicit through the Borel 
re-summation of perturbative expansions about
the classical solutions.  
For each classical solution,  there is a loop expansion of
the partition function of the form 
\bg
Z_i \equiv \sum_{n}(\sqrt{ {\pi\over   \hbar S_{i}^{\prime\prime} } })^N  
e^{-{1\over \hbar}S_i  }
	c_{n,i} \hbar^n 
\label{brel}
\ed
where $i$ labels  the classical solution, and $N$ is the number of 
degrees of freedom. 
This expansion is asymptotic,  but its Borel transform
\bg 
B_i  = \sum_{n} {1\over n!} c_{n,i} t^n
\label{BRL}
\ed 
is believed to have a a non-zero radius of convergence under very general 
conditions
in field theory.
Term by term,  the inverse of the Borel transform is given by  
\bg
Z_i = (\sqrt{ {\pi\over   \hbar S_{i}^{\prime\prime} } })^N 
e^{-{S_i\over \hbar}  }
    \int _0 ^{\infty} e^{{ -t\over \hbar }} B_i(t)
\label{brig}
\ed
However this is often ill defined,  as $B_i(t)$ may have singularities on 
the positive 
real axis due to instantons and renormalons.   In some simple cases one can 
prove
that as long as the contour of the
$t$ integration is closed,   or begins at $t=0$  and ends at $Re(t) = 
+\infty$ avoiding
all singularities, then one
obtains an exact,  and usually  exotic solution of the Schwinger-Dyson 
equations
satisfying the action principle\footnote{Note that sometimes the 
integral over $t$ for an open contour 
is not always convergent\cite{thooft}.  We will not address this  
this subtlety 
here.}.  Sometimes this can be simply proven by analytic 
continuation of the couplings from another  situation in which the couplings
are real and there are no 
singularities on
the positive real axis.  In Appendix A, we present a more direct and 
general proof
for an arbitrary theory with one degree of freedom.  Note that as long 
as the action has
no flat directions, and the perturbative parameter is not infinite for 
any of the classical 
solutions,  then the number of 
classical solutions is equal to the number of independent solutions of 
the Schwinger
Dyson equations.  As a simple example consider the theory 
$S= {g\over 4} \phi^4 +{\mu\over 2}\phi^2$.  Perturbation theory is an 
expansion in 
${g\over\mu^2}$.  For non-zero $\mu$,  there are three classical 
solutions and three 
independent solutions of the Schwinger-Dyson equations.   However at 
$\mu=0$  there
is only one classical solution $\phi = 0$,  but still three 
independent solutions of 
the Schwinger-Dyson equations.  We shall address such issues as 
flat directions and 
renormalons in the context of boundary conditions for the 
Schwinger-Dyson equations 
in a later work.
Assuming that perturbative parameter for any classical solution is finite, 
then an arbitrary solution of the
Schwinger-Dyson equations consists of a linear combination of 
solutions generated from 
the classical solutions.  Which solution is generated by a particular 
classical solution is, of course, somewhat arbitrary.  For instance in 
the $\phi^4$ theory
the resumed perturbation expansion about the classical solution 
$\phi=\sqrt{-\mu\over g}$
for $\mu>0$ yields either $Z^+$ or $Z^-$,  depending on how one avoids the 
singularity in $B(t)$  due to the neighboring classical solution at  
$t= S(\phi = 0) = 0$. 
	
The point we now wish to make is that an exotic solution of the 
Schwinger-Dyson
equations given by $Z= c_i Z_i$,  where the $Z_i$  are generated from 
classical solutions,
is a generalized form of theta vacuum.  The $c_i$ play the role of  
theta parameters.  
If the theory has a space-time with a boundary,  then these theta 
vacua are simply related to conventional theta vacua.  In the usual 
approach a particular
theta vacuum is selected by adjusting a surface term in the action and 
integrating over 
real fields.  The surface term term effects the Schwinger-Dyson 
equations only at the 
space-time boundary,  so in the infinite volume limit its role 
is only to set a boundary
condition.  It does this by putting a different weight on the 
contributions to $Z$ 
coming from fluctuations about different real classical solutions,  which
when Borel re-summed correspond to different complex contours.   
Thus, in the thermodynamic limit,
it seems clear that the real integration with a surface term in the 
action is equivalent to an exotic solution  
$Z= c_i Z_i$,  although we shall not prove it with any rigor. 

There is a class of boundary conditions which we have not yet discussed.
These boundary conditions arise because on a lattice with a boundary,  
the measure at the boundary is completely arbitrary.  It does not have
to correspond to a surface term in the action which puts different weights 
on the expansion about different instantons,  or equivalently to some choice
of contours on a lattice without a boundary.   For example 
in a field theory any matrix element of the form
\bg
<\psi| Te^{\int d^Dx J(x)\hat{\phi(x)} } |\psi^{\prime}>  
\label{melem}
\ed
where the wave-functions $\psi$ and $\psi^{\prime}$ are independent of $J(x)$,
satisfies the Schwinger-Dyson
equations.   These solutions are obtainable from the real contour path 
integration 
with the wave functions setting the measure at the time 
boundaries.  For instance in quantum mechanics one would write 
\bg
Z=\int D\phi(t) \psi(\phi(+T))\psi^{\prime}(\phi(-T))
e^{-S[\phi(t)] + \int dt J(t)x(t)}
\label{buk}
\ed
Such boundary conditions are necessary to obtain generating
functionals in the excited states,  which can not be found from the 
generalized theta vacua discussed above.  
By considering a free lattice field theory, it is easy to see that 
excited state solutions may not be obtained from 
exotic contours with periodic boundary conditions. 
Consider the Schwinger Dyson equations
\bg
(\sum_y \del^2_{x,y}
{\del\over\del J_y} + \mu  {\del\over\del J_x} -J_x) Z[J] =0.
\label{fre}
\ed
In Euclidean space on a periodic lattice $\del^2$ is invertible and these
equations have a unique solution up to a 
normalization.  Unsurprisingly, there are no inequivalent contours in the 
integral
representation.  Therefore to
get the solutions which in the continuum limit correspond to generating 
functionals in all the excited
states,  one must consider the same equations with a temporal boundary. 
However in Euclidean space,  the excited state Green's functions are distinct
from the vacuum Green's functions in that they do not
satisfy cluster properties.  For example $<1|T(\hat O(t) \hat O(0))|1>$
behaves like $|<1|\hat O(0)|0>|^2\exp{ (E_1 - E_0)t }$  at large
$t$.   In this paper we will confine our discussion to generating functionals
in the vacuum sector,  for which 
the set of exotic contours on a periodic lattice is a complete set
of boundary conditions.  

Note that different measures at a spatial boundary  are not always 
available to set the boundary conditions. An example is given
by the one matrix model.
There are symmetry breaking solutions
of the ${g\over 4}Tr M^4 + \dots$ matrix model with real positive $g$  
which are not 
obtainable 
from the real contour by any deformation of the 
action\cite{CI}\cite{zackparis}\cite{nextmmpaper}.  However all the solutions 
may be obtained using
the complete set of exotic 
contours \cite{zackparis}\cite{nextmmpaper}
\cite{davide}\cite{kitaev}.
 
\section{Solution Set in the Thermodynamic Limit}

According to the previous arguments,  any theory with a finite number of 
degrees of freedom
will have multiple theta parameters which continuously relate all the 
solutions of the
Schwinger-Dyson equations.  In general the number of lattice
theta parameters grows exponentially
with the number of degrees of freedom.  
Thus it appears that there is an intractable number of solutions.  
Fortunately,  
not all of 
these solutions can survive,  or are distinct, in the thermodynamic limit.  
This limit exists if the free energy is extensive,  
meaning that
the partition function $Z$ behaves like
$e^{-{\cal N}f}$ as the number of degrees of freedom or the volume of the 
system, 
${\cal N}$, goes to infinity.
In the thermodynamic limit, a linear combination of two solutions for which 
the free 
energy is extensive, 
$Z=ae^{-{\cal N}f_1}+be^{-{\cal N}f_2}$,   is either equivalent 
to the solution with the  lowest free energy density, or does not have a 
thermodynamic limit if the real parts of 
$f_1$ and $f_2$ are equal.  Thus there are really only two solutions rather 
than a 
continuum of 
solutions labeled by the  parameter ${a\over b}$. 
In a Euclidean field theory the absence of a thermodynamic limit is reflected
in the failure of 
the vacuum Greens functions to cluster.  In an $N x N$ matrix model it 
is reflected in the 
absence of a genus expansion in ${1\over N}$.

The absence of this continuum of solutions is the reason that there is no
spontaneous symmetry breaking in continuum quantum mechanics,   even though 
on a finite discrete time lattice,
the Schwinger-Dyson equations have real symmetry breaking solutions.  
Consider the partition function of the anharmonic oscillator, 
$V(x) = {g\over 4}x^4 + {\mu\over 2}x^2$  with $\mu<0$,
with periodic boundary
conditions in time,  and a period given by $\beta$.  
There is symmetry breaking
at every order in the perturbative expansion about $x=\sqrt{-\mu\over g}$. 
Due to tunneling into the other well,  the Borel transform of this 
expansion has a singularity on the
positive real axis.  By avoiding this singularity one obtains an exact but 
complex
symmetry breaking solution of the Euclidean
Schwinger Dyson equations.  Furthermore since the free energy of this 
solution is
extensive in the perturbative expansion,  its Borel re-summation is also 
extensive.  
In other words, for large $\beta$ this solution is of the form  
$Z=e^{-\beta E_0}$,  where $E_0$ is complex.
Since the Schwinger Dyson equations are linear,  one can add other 
solutions in many
ways to cancel the imaginary parts.   For example $Z+Z^*$ is a real, 
symmetry breaking
solution of
the Schwinger Dyson equation.  However the real parts of the free energy
densities associated with
$Z$ and $Z^*$ are equal.  Therefore the sum is not extensive,   
and the Euclidean greens functions 
do not satisfy cluster properties.  In some cases it is possible
to combine solutions with imaginary parts and still obtain an
extensive solution. For example, it has been 
shown\cite{zinnjustin}
that if one considers
the instanton gas sum for the anharmonic oscillator,  then the various complex 
parts arising from the attractive interactions of the instantons,  
and the Borel 
re-summations in different instanton sectors,  all cancel,  yielding a 
real solution
which is extensive and symmetric.  In this case
a real and extensive solution is obtained by adding
an {\it infinite} number of complex extensive solutions.
There may be several ways to perform such infinite sums. 
In the quantum mechanical sine-gordon model,  the different ways of doing 
so yield
a periodic ``theta''  band,  and in a field theory yield the conventional
theta vacua.

Presumably in the broken phase of a $\phi^4$ field theory,  there are real 
symmetry 
breaking sums over contours for which the thermodynamic limit exists,  
whereas for the integration over real fields there is no such limit.  
In this case one can still use the integration over real fields if one adds a 
symmetry breaking term at the space-time  boundary,  or a global symmetry 
breaking term which is removed
after taking the thermodynamic limit.   However, we wish to emphasize 
again that 
the full set of
complex integration contours is sometimes necessary to account for all 
the solutions.
This is especially clear in certain real double scaling solutions of matrix 
models\cite{CI}\cite{zackparis}\cite{nextmmpaper}\cite{davide},
and also in false vacuum solutions of field theories. 
In the semi-classical limit, small perturbations of the action  which are
removed upon taking the thermodynamic limit tend to pick out quantum states 
which are 
are peaked about a particular choice among degenerate classical global 
minima.  
However, when there are 
non-degenerate minima,  there are generally complex false vacuum solutions 
corresponding to the 
sub-dominant minima, and these may not be obtained from the real contour 
integration.
As one evolves these solutions through a first order phase transition,  
they may be
obtained from a sum over contours which is real. This discontinuous change 
in the 
list of boundary conditions which lead to a thermodynamic limit is typical 
of crossing
a phase boundary. 
As we shall argue next, 
certain features of critical phenomena are also simply related to the 
existence 
of the 
generalized set of boundary conditions of the Schwinger-Dyson equations.

In a one-matrix model is not difficult to see that in the thermodynamic 
limit,  extensive 
solutions coalesce in such a way that there are either discrete sets of 
solutions or a
countable number of theta parameters\cite{zackparis}\cite{nextmmpaper}.  
It is difficult to prove this in a general
lattice field theory.  Instead we will assume that the solution set 
collapses in a way
which depends on the coupling constants,
and argue that this leads to expected physical consequences.  One such 
consequence is
the appearance of phase boundaries.  In a finite system,  as one 
makes large changes
in the couplings one must change contours to maintain convergence.  
In the thermodynamic
limit,  one may have also have to make   
changes in the boundary conditions so that this limit continues to exist,
and so that the solutions vary smoothly
in the space of coupling constants. 
For a finite number of degrees of freedom
this additional change in the boundary conditions (the sum over integral 
contours)  violates the Schwinger
action principle,  as discussed in section II.  
However in the thermodynamic limit it is possible to make this change 
in the boundary 
conditions and satisfy the
action principle if 
solutions with different boundary conditions coalesce.  
Let us assume that for every  
solution in the thermodynamic limit,  there is a 
set of associated boundary conditions.  Then a sufficient condition 
for two solutions at 
infinitesimally different
values of a coupling to be related by the action 
principle is that these sets have
some intersection. 
In order that a solution vary smoothly in the the thermodynamic limit, 
certain boundary conditions may become disallowed as one varies the couplings.
However since there are generally large
sets of boundary conditions associated with a solution,  one can find boundary
conditions which may be held fixed at least for small variations of the 
couplings.  For large variations of the couplings one may be forced to change
the boundary condition,  but this does not violate the action principle
the way it would for a finite number of degrees of freedom,  in which case
different boundary conditions correspond to distinct solutions.  
Thus by making a large loop in
the complex plane of some coupling constant,  one can return 
to a different boundary 
solution than the one with which one started\footnote{Starting from a 
physical solution, 
such analytic continuations 
may lead to solutions which do not satisfy physical constraints,
or to other physical solutions.}.  
Therefore because of the collapse of the solution set in the thermodynamic 
limit,  
phase boundaries can appear in the form of branch points in certain
coupling constants.
If one were to keep the boundary conditions fixed globally, then there would
have to be a barrier to smooth analytic continuation,  which according
to the standard lore is formed by the accumulation
of Lee-Yang zeroes\cite{Lee-Yang} of the partition function.   

To clarify the arguments above,  we describe 
a simple zero dimensional analogue,  
in which the solution set shrinks accompanied by the accumulation of Lee-Yang
zeroes and the appearance of a phase boundary.  
Consider a solution of the
Schwinger-Dyson equation for the
action $S(\phi)= {g\over 4}\phi^4 + {\lambda\over 3}\phi^3+\dots$.  
If the action principle
is satisfied,  then
the generating function $Z$ is analytic in $\lambda$ (except at infinity) 
as long as $g$ 
does not vanish.  There is
no need to rotate contours to maintain convergence as one changes the 
phase of $\lambda$.  
However if one sets $g$ to zero,  then the order of the Schwinger-Dyson 
equation
$(g\del_J^3+\lambda\del_J^2 + \mu\del J - J)Z=0$ drops by one,  and the space 
of solutions 
becomes smaller.  In this sense, the limit $g\rightarrow 0$ is analogous to 
the 
thermodynamic limit. 
Different boundary conditions give different asymptotic behavior for
small  $g$.  Some solutions do not exist in the $g\rightarrow 0^+$  
limit,  while others
coalesce.  The sets of boundary conditions which survive and coalesce 
in this limit depends on
the value of $\lambda$.   Consider the limit 
$g\rightarrow 0$ 
with $g$ real and positive.  The sets of equivalent convergent integration 
contours for $g \ne 0$ differ from those for $g=0$ (fig.3). 
The intersection of these sets determines how solutions of the theory 
behave as $g$ 
is set to zero.   If the set of equivalent contours associated with a solution 
for $g\ne 0$ has no intersection
with the set of equivalent contours associated with any solution for $g =0$,
then there is no $g\rightarrow 0$ limit.  For instance the integral over the
real axis behaves as $Z\simeq exp({\lambda^4 \over 12 g^3})$ as 
$g\rightarrow 0^+$ with $\lambda$ real and positive.     
For certain values of 
$\lambda$  there are solutions for which the intersection 
consists of contours which can be closed in the $\lambda\phi^3$ theory,  
giving 
$Z \rightarrow 0$ as $g\rightarrow 0$.  Therefore at these values of $\lambda$,
there are many solutions which coalesce as $g\rightarrow 0^+$ due to the 
linearity of the Schwinger-Dyson equation.     
At $g=0$ maintaining convergence forces one to rotate the 
contours as one rotates the phase of 
$\lambda$ by $2\pi$,  and the solutions become triply sheeted in $\lambda$ 
with a 
phase boundary at 
$\lambda = 0$. 
As one changes the phase of $\lambda$ at small but non-zero $g$, 
small violations of the action principle allow one to move among solutions
in a way which at $g=0$ corresponds to analytic continuation among 
Riemann sheets in the complex $\lambda$ plane.  
This amounts to altering the boundary conditions to prevent Stokes
phenomena from occuring  on the positive real $g$ axis as one changes the 
phase
of $\lambda$.  
If one were to keep the boundary conditions (contours) 
fixed at non-zero $g$,  then $Z$ 
would be a single valued function of 
$\lambda$
and there would have to be some obstruction to analytic continuation 
of $\lambda$ in a small $g$ expansion.  One way this obstruction
can appear 
is by the accumulation of Lee-Yang zeroes of $Z$ along a Stokes line
in the complex $\lambda$ plane as $g\rightarrow 0^+$.
At small but non-zero $g$ one can change the location of the Stokes line by
changing $\lambda$ or by changing the boundary conditions.  
By the appropriate change in the boundary conditions as one varies $\lambda$,
one can prevent the Stokes line from crossing the positive real $g$ axis. 
The reason it is possible to make such changes 
in the boundary conditions with only small violations of the action principle
is that solutions with different boundary conditions coalesce as 
$g \rightarrow 0$.   

For the example we have given,  it is difficult to see if the accumulation of 
zeroes explicitly.  However there is an even  
simpler example.  Consider 
starting with the theory $S = {\lambda\over 3}\phi^3 +{\mu\over 2} \phi^2$,  
and take the limit
$\lambda\rightarrow 0^+$.  In this case the partition function is 
exactly given by
\bg
Z(\lambda,\mu) = \lambda^{-{1\over 3}} e^{ -{\mu^3\over 12 \lambda^2} }
       (\alpha Ai( {\mu^2\over 4 \lambda^{4\over 3}  } )+ 
	\beta Bi( {\mu^2\over 4 \lambda^{4\over 3}  } ) )
\label{airy}
\ed
where $Ai(x)$ and $Bi$ are the two independent Airy 
functions.  The combinations
which survive in the $\lambda \rightarrow 0^+$ limit approach the free field 
solution
\bg
Z(0,\mu) =\sqrt{ {\pi\over \mu} }
\label{fuge}
\ed
At finite $\lambda$ there is no branch point in $\mu$,  although the  
$\sqrt{\mu}$ analytic
structure does appear at any finite order in a large $\lambda$ expansion.  
One way
this can occur is if $Z$ has an infinite number of zeroes in $\lambda$
which become dense at infinity.  As $\lambda \rightarrow 0^+$ we expect
these zeroes to accumulate also at finite values of $\lambda$. 
Because of the ${\mu^2\over 4 \lambda^{4\over 3} }$ in the argument of the 
airy function,  it is clear that this is indeed the case,  with 
zeroes accumulating on the negative real axis\footnote{Of course this is 
a somewhat trivial example of a phase boundary,  
since the difference between the two Riemann sheets 
in $\mu$ is only a 
normalization.  A $2\pi$ change in the phase of $\mu$ leads to a 
rotation of the contour 
by $\pi$,  which changes only the sign of $Z$ and not any of the 
Green's functions,  which
have a pole rather than a branch point at $\mu=0$.}.  As one rotates the 
phase of $\mu$,  the lines of zeroes rotate in the complex $\lambda$
plane,  unless one simultaneously changes the boundary conditions, namely
$\alpha$ and $\beta$.  

It is quite generally true that the collapse of the solution set of
the Schwinger-Dyson equations is intimately connected to the accumulation
of Lee-Yang zeroes and the appearance of phase boundaries, whether this 
collapse
is brought about by tuning parameters to special values,  or by taking 
the thermodynamic limit.  The collapse of the solution set forces one
to change the boundary conditions while making large variations of the 
couplings in order that solutions depend smoothly
on the coupling constants.  At the same time the collapse allows one to 
make these changes without violating the action principle.   
Consequently it becomes possible to have solutions with branch point 
singularities, and  
critical exponents are determined by  
the manner in which the solution set collapses.  
It can also be shown in certain matrix models that analytic continuation of
a solution with a real double scaling limit and very conventional boundary
conditions (real axis integration) leads to other solutions which also
have a real double scaling limit,  but very unconventional 
boundary conditions\cite{nextmmpaper}.   
These unconventional boundary conditions
can not be written by assigning a single equivalent  sum over integral 
contours to every eigenvalue. In other words, these solutions may not be 
written in the manner of figure 2(a).

\begin{figure}

\centerline{
\epsfxsize=11cm
\epsfbox{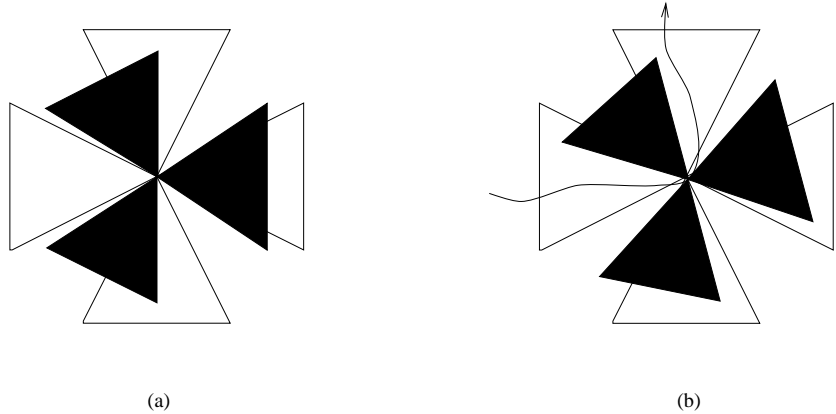}  }

\bigskip
\caption{(a) The allowed angular domains of integration as  
$|\phi| \rightarrow \infty$ are drawn for $S = g \phi^4 + \dots$ in white 
and for 
$S = \lambda \phi^3 + \dots$ in black,  where $g$ and $\lambda$ are real 
and positive.
The intersection of these domains determines the behavior of solutions as 
$g\rightarrow 0$.  (b) is the same diagram drawn for a complex $\lambda$. 
The solution associated with the depicted contour has a vanishing partition
function as $g\rightarrow 0$, keeping the phase of $g$ fixed.}
\label{fig3}
\end{figure}

We have argued above that the collapse of the solution set in various limits
forces one
to make changes in the boundary conditions when one varies the couplings, 
and that these changes may be consistent with the action principle.
While this is clearly true in the zero dimensional examples, 
the arguments that it is true in a thermodynamic limit were only 
heuristic.   
Below we show that the action principle follows from
the Schwinger Dyson equations in the thermodynamic limit.  
For a finite number of degrees of freedom the action principle and
the Schwinger Dyson equations are
independent. The manner in which a solution changes as one varies a 
coupling is arbitrary for a finite number of degrees of freedom,  
since at any fixed coupling 
all the solutions are continuously related. 
However let us assume that in the thermodynamic limit there are only
discrete sets of solutions of the Schwinger-Dyson equations at fixed 
values of the
couplings,  and none that are continuously related. 
Let us also assume that solutions
vary smoothly as one varies the couplings,  with exceptional points  
corresponding
to phase boundaries.  
We shall define the couplings $g_{\alpha}$ so that they appear 
holomorphically in the action 
and the Schwinger-Dyson equations.  The solutions will then be holomorphic 
functions of the couplings\footnote{While analyticity holds in the 
thermodynamic limit,  it does not imply that analytic continuation of a 
physical solution around a phase boundary yields a solution which is still
physical \cite{turbiner}.}. Therefore
some first order differential equations of the form 
\bg
\hat O _{\alpha} Z \equiv 
( {\del\over\del g_{\alpha} } - \hat K_{\alpha} )Z[g,J]=0.
\label{opdef}
\ed 
are automatically induced when the solution set becomes 
discrete\footnote{If the solutions in the
thermodynamic limit are not discrete,  but have a countable number 
of continuous theta parameters,  then
the Schwinger action principle is not induced,  but may certainly be 
consistently imposed.}.
More precisely we should say that there is a discrete set of solutions for
$f = {1\over {\cal N}}ln Z$ 
as ${\cal N} \rightarrow \infty$,  and it is this quantity 
which satisfies some set of first order 
differential equations in the couplings. 
The solutions for $Z$ associated with one of the discrete solutions for $f$
may be written as  $Z = e^{-{\cal N}f} + \delta$  where $\delta$
is subdominant in the large ${\cal N}$ limit.  There are generally many 
possibilities for $\delta$.  Therefore strictly speaking the $Z$'s associated
with a particular $f$ in the ${\cal N}\rightarrow\infty$ limit 
will satisfy an equation of the form 
(\ref{opdef}) up to some arbitrary terms subdominant to $e^{-{\cal N}f}$.
These arbitrary terms depend on how one chooses $\delta$, given $f$.

It remains to show that these equations correspond to the action principle.
We shall assume $\hat K$ is a linear operator,  since a generic  nonlinear term
would be either dominant or sub-dominant to 
${\del\over\del g_{\alpha} }Z$  in the thermodynamic limit.  For instance
$Z^2 \sim  e^{-2{\cal N}f}$  while ${\del\over\del g_{\alpha} }Z \sim 
e^{-{\cal N}f}$,  
as ${\cal N} \rightarrow \infty$.
Thus we have 
\bg
\hat O _{\alpha} = 
{\del\over\del g_{\alpha} } - K[J_i, {\del\over\del J_k}, g_{\beta}] 
\label{opde}
\ed
Let us denote the Schwinger-Dyson
equations by $\hat L_i Z=0$,  where the index $i$ labels the degrees of 
freedom. 
Since $\hat O _{\alpha}$ is linear,  
$[\hat O_{\alpha},\hat L_i] Z = 0$,  up to terms subdominant to
$e^{-{\cal N}f}$.  
We decompose $\hat O_{\alpha}$ into a piece which commutes with 
all the $\hat L_i$ and a piece which does not;
\bg
\hat O_{\alpha}= \hat H_{\alpha} + \hat M_{\alpha}
\label{decomp}
\ed
where 
\bg
[\hat H_{\alpha}, \hat L_i] = 0
\label{commut}
\ed
and the derivative with respect to the coupling is included in 
$\hat H_{\alpha}$;
\bg
\hat H_{\alpha} = {\del\over\del g_{\alpha} } -\dots
\label{derv}
\ed
With this decomposition one has
\bg
\lim_{{\cal N}\rightarrow\infty} e^{+{\cal N} f}[\hat O_{\alpha},\hat L_i] Z 
=\lim_{{\cal N}\rightarrow\infty} e^{+{\cal N} f}\hat L_i \hat M_{\alpha} Z=0
\label{scromp}
\ed
Therefore, $M_{\alpha} Z$ may be written as the sum of any solution 
of the Schwinger Dyson equation $Z^{\prime}$ and a term subdominant to 
$e^{-{\cal N}f}$.  However for the 
equation $\hat O_{\alpha} Z=0$  to make sense 
in the thermodynamic limit,  $Z^{\prime}$ can not be dominant with respect to
$Z$ in the ${\cal N}\rightarrow\infty$ limit. 
Therefore $Z$ must be an an approximate eigenfunction of $\hat M_{\alpha}$; 
\bg
\hat M_{\alpha}Z = \gamma(g_k) Z  
\label{fnk}
\ed
up to terms subdominant to $e^{-{\cal N}f}$,
which means that   
\bg
(\hat H_{\alpha} + \gamma(g_k))Z = 0,
\label{cool}
\ed
up to terms subdominant to $e^{-{\cal N}f}$.

The set of operators $\hat H_{\alpha}$ of the form ${\del\over \del 
g_{\alpha} } -\dots$
which commute with $\hat L_i$  are in fact the operators
associated with the action principle.  To see this, consider an 
arbitrary 
zero dimensional theory with the action 
$S=\sum_{\alpha} g_{\alpha} f_{\alpha}(\phi)$.   
The operator associated with the Schwinger Dyson equations
is 
\bg
\hat L = \sum_{\alpha} g_{\alpha} f_{\alpha}^{\prime}(\del_J)-J  
\label{qrk}
\ed
and the operators associated with the action principle are
\bg
{\del\over\del g_{\beta}} + f_{\beta}(\del_J)
\label{skk}
\ed
These operators commute.
For instance with $S= {g\over 4}\phi^4$ one finds that
\bg
[{\del\over\del g} + {1\over 4} \del_J^4,  g\del_J^3 - J] = 0 
\label{e}
\ed
This commutation relation generalizes in an obvious way to theories with 
more than one degree of 
freedom.  Note that one can add a term which is constant in the source J,  
but dependent on the couplings, to   
any of the operators associated with the action principle without changing the
commutation relations with $\hat L$.  
There are some restrictions on what this additional term may be,  since the 
change
$\hat H_{\alpha} \rightarrow \hat H_{\alpha}+ \gamma_{\alpha}(g_1,g_2,...)$
leads to 
\bg
[\hat H_{\alpha},\hat H_{\beta}] = 
{\del\over g_{\alpha} } \gamma_{\beta} - {\del\over g_{\beta} } \gamma_{\alpha}
\label{comt}
\ed
which must vanish if $\hat H_{\alpha} Z = 0$  
The addition of a term $\gamma_{\alpha}$ satisfying this constraint 
has no effect on
the action principle, since $\gamma$  may be 
absorbed by a change in the normalization of $Z$ 
\bg
Z \rightarrow e^{-\int^g dg^{\prime}_{\alpha} \gamma_{\alpha} (g^{\prime} ) } Z
\label{ren}
\ed
If the Schwinger action principle is written in terms of normalized 
Green's functions,  the
$\gamma$ term does not appear at all.  Thus the action principle follows 
from the Schwinger
Dyson  equations in the thermodynamic limit.

\section{Analytic Structure of the Effective Potential}

In this section we discuss the effect of the collapse of the solution set 
on the 
effective potential. 
In general,  for a finite number of degrees of freedom,  there is
a continuous set of effective potentials,  with extrema associated with a  
single or a 
small discrete set of solutions of the Schwinger-Dyson equations.
Recall that the effective potential 
may be defined as follows.  Take for example the $\phi^3$ theory with
action $S={\lambda\over 3}\phi^3 + {\mu\over 2}\phi^2$.  
In terms of $\phi(J) = {d\over dJ}\ln Z(J)$  the Schwinger-Dyson equations 
may be 
written as 
\bg
J=\lambda\phi^2(J) + \mu\phi(J)  + \lambda\del_J\phi(J)
\label{fijh}
\ed
which has a one parameter class of solutions.  Given a solution $\phi(J)$ the 
associated effective potential $\Gamma(\phi)$ is then defined 
by the solving the equation
\bg
J= \Gamma^{\prime}(\phi).
\label{eff}
\ed
This is an algebraic equation rather than a differential one, 
and in general $\Gamma(\phi)$ can not yield the full continuous one parameter 
class of solutions,
but only a discrete subset. 
There is an extremum of $\Gamma(\phi)$ associated with each 
branch of $\phi(J)$ in the complex $J$ plane.
However in zero dimensions $Z$ is analytic in $J$,  and $\phi(J)$ will 
in general 
have an infinite
tower of poles in $J$.  In our example these poles arise from the infinite 
tower
of zeroes of 
\bg
Z(J)= \lambda^{-{1\over 3}}e^{ -{\mu^3\over 12 \lambda^2} + 
{\mu\over 2\lambda} J }
       Ai( {\mu^2\over 4 \lambda^{4\over 3}  } + {J\over\lambda^{1\over 3} } )
\label{aigain}
\ed
in the complex $J$ plane.
Since there are an infinite number of poles but no branch points in J,  
the analytic 
structure of $\Gamma (\phi)$ is somewhat odd.  Because of the infinite 
number of poles,
$\phi=\infty$  corresponds to an infinite number of discrete values of 
$J = \Gamma^{\prime}(\phi)$,  so $\Gamma (\phi)$  has an infinite number 
of Riemann
sheets in the complex $\phi$ plane.  Also, since $\phi(J)$ has no branch 
points in $J$,
there is a unique value of $\phi$ at which  $J= \Gamma^{\prime}(\phi) = 0$. 
Therefore there is a different effective potential for every solution.  
In general one must Legendre transform with respect to a variable in which the 
partition function is multiply sheeted in order to get an effective 
potential with
multiple extrema.
Note that at any finite order in the loop expansion,  one gets an 
effective potential 
which appears to have more than one extremum.   
In zero dimensions this is a spurious feature of the
loop expansion,  which may be recast as an asymptotic expansion for large J.
For instance, in the $\lambda \phi^3$ example,  the loop expansion 
corresponds to writing
\bg
J=\lambda\phi^2(J) + \mu\phi(J)  +  \hbar \lambda\del_J\phi(J)
\label{semic}
\ed
Perturbing in powers of $h$,  and setting $h=1$ gives an expansion of the form
\bg
\phi(J) = a(J - J_c)^{1\over 2} + b + c(J - J_c)^{-1\over 2} + \dots
\label{expj}
\ed
where $a,b,c \dots$ and $J_c$ are functions of $\mu$ and $\lambda$.
At any finite order there is a square root branch point at $J = J_c$.
In a thermodynamic limit however,  it is possible that the analytic 
structure apparent 
in the loop expansion also corresponds to the analytic structure of the 
exact solution.
In a limit in which the Lee-Yang zeroes of $Z(J)$  accumulate to give
a multiply sheeted solution, 
a single effective potential will describe several solutions.
Furthermore,  if solutions coalesce as they are expected too,  then so will 
the associated
effective potentials.  There will not be 
a continuous class of effective potentials, except in cases in which 
a theta parameter survives.  

It is generally difficult to find the boundary conditions associated 
with a certain solution.
However for the sake of illustration,  let us consider the 
three extrema of the effective potential
in the broken phase of $\phi^4$ field theory, and make an educated guess 
as to the boundary conditions associated with each of these extrema.  The 
two symmetry 
breaking extrema are probably obtained in the continuum limit of the 
lattice theory using
contours such as 
\bg
\prod_x \int_{\Gamma^-}  d\phi_x + \prod_x \int_{\Gamma^+} d\phi_x  
\label{produ}
\ed
or
\bg
\prod_x \int_{\Gamma^- +\Gamma^+}  d\phi_x
\label{produ2}
\ed
where $\Gamma^+$ and $\Gamma^-$  are as defined previously in (\ref{gammas}). 

The effective potential has another extremum at $\phi = 0$,  
corresponding to an 
unphysical solution which survives in the thermodynamic limit.   
It is somewhat more difficult 
to guess which boundary conditions  this solution  corresponds to. 
We need another boundary condition which has a 
thermodynamic limit for $\mu<0$  and which
is symmetric under $\phi\rightarrow -\phi$ 
The lattice theory has a simple property
which helps to find this boundary condition.
The lattice action is
\bg
S= {1\over2}\sum_{x,y,\hat{\mu}} \delta_{x,y + \hat{\mu} } \phi_x \phi_y + 
   \sum_x {g\over 4} \phi^4 + ( {\mu\over 2} - D ) \phi^2
\label{latac}
\ed
Where $D$ is the dimension.
Consider an arbitrary boundary condition and rotate the contours 
by ${+\pi\over 2}$ 
and ${-\pi\over 2}$ at 
alternate lattice sites.  This is equivalent to keeping the contours 
fixed,  but flipping 
the sign of the $\phi^2$ term and rotating the phase of the source 
term $J_x$  by 
$i$ and $-i$  at alternate lattice sites.  The interaction and hopping 
terms are
invariant.  When the sign of the $\phi^2$ term is 
flipped $\mu\rightarrow  -\mu+2D$,  so this mapping relates the broken 
and unbroken phases.
Note that this is not a duality in the usual sense,
since the boundary conditions are changed as well as the parameters of 
the action.
If for $\mu<0$ we choose contours of integration which get mapped to the
real contour,  then via this mapping the solution is  
equivalent to the one obtained from the real contour in an unbroken phase. 
The only difference is in factors of $-1$ for certain Green's functions,  
due to the rotation
of the phase of the source term. 
We conjecture this solution corresponds to the local maximum of the 
effective potential.
It would certainly not be surprising,  since for $\mu<0$ in zero 
dimensions the
local maximum of the effective potential at any finite order in the in the 
loop expansion 
corresponds to the solution obtained by integrating along the imaginary axis.

\section{Conclusions}

The phase structure of bosonic field theories and 
matrix models appears to have a very natural interpretation 
in terms of the multiple boundary conditions of Schwinger-Dyson equations. 
In the thermodynamic limit,  phenomena such as the accumulation of Lee-Yang
zeroes and the collapse of the solution set are intimately related.
There are several interesting questions which we are pursuing, 
of which we will list only a 
few.  It is not known whether or how these ideas may be generalized 
to fermionic 
theories.  The Schwinger-Dyson equations of a fermionic theory may be
written,  as in the bosonic case,  as a set of recursion relations among
Green's functions.  However for fermions on a finite lattice there are  
countable number
of Green's functions.  Therefore these recursion relations must truncate.
Because of this truncation,  there is not a large set of solutions
when the number of degrees of freedom is finite.
This may have interesting consequences for the large field behavior of
the action in any bosonized version of the lattice theory.    
Another question concerns the relation between the
various boundary conditions and the phases of a theory with a nonlocal 
order parameter.
As yet we have not been able to construct such a relation,  though we 
conjecture that it exists.  We are also pursuing the question of 
precisely how
renormalons are related to the multiple boundary conditions of the
Schwinger-Dyson equations.  In the simple examples considered in this
paper the perturbative parameter is finite,  and there is a one to one
relation between instantons and a basis set of solutions of the
Schwinger-Dyson equations,  with the only singularities in the Borel 
transform arising from instantons.   When there are renormalon 
singularities as well,  then there are more inequivalent ways to avoid 
singularities when integrating over the Borel variable. This leads to 
the conjecture that there are more boundary conditions for Schwinger-Dyson 
equations than instanton counting would indicate.  

\medskip
\noindent{\bf Acknowledgment}
\medskip

The authors wish to thank Stephen Hahn  and
Paul Mende for useful discussions.  
S. Garcia has recieved support in part from
NSF Grant ASC-9211072 and DOE Grant DE-FG09-91-ER-40588 - Task D.
G. Guralnik is supported in
part by funds provided by the U.S. Department of Energy (D.O.E.)
under cooperative agreement \#DF-FC02-94ER40818 (M.I.T.) and by 
Grant DE-FG09-91-ER-40588-Task D. (Brown)\ 
He would like to thank John Negele for hospitality at 
the CTP.   Z. Guralnik recieves support from NSF grant PHY-90-21984. 

\section{Appendix I}

\subsection{Exact solutions of zero dimensional $\phi^4$ theory}
We have discovered, not surprisingly, that considerable effort has 
been devoted to solving the zero dimensional $\phi^4$
theory)\cite{scarpetta}, \cite{caianello},\cite{caianello2},
\cite{symbreak},\cite{klauder},\cite{cooper}. 
Much of what we present in this section has been known in one form or another.
It is the intent of our presentation to bring new clarity and 
completeness to the solutions
of this model in a form consistent with the point of view of this paper.
Here we construct the exact Green's functions for the full solution set of the
zero dimensional theory 
with action $S={g\over 4} \phi^4 + {\mu\over 2} \phi^2$.
The normalized disconnected Green's functions $G_n$ are defined by the Taylor 
expansion of the generating function,
\bg
{Z(J)\over Z(0)} = {1\over n!} G_n J^n
\label{tay}
\ed
The Schwinger-Dyson equations may be written as the recursion relation
\bg
gG_{n+3} + \mu G_{n+1} - n G_{n-1} = 0
\label{rec}
\ed
The initial conditions for this recursion relation are the one and two 
point function. 
The Schwinger action principle may be written as 
\bg
{\del G_n \over \del g} - {1\over 4} G_4 G_n + {1 \over 4} G_{n+4} = 0
\label{sdys1}
\ed
and
\bg
{\del G_n \over \del \mu} - {1\over 2} G_2 G_n + {1 \over 2} G_{n+2} = 0
\label{sdys2}
\ed
Since the Schwinger-Dyson equations allow one to write all Green's functions
in terms of $G_1$ and $G_2$,
the action principle may be rewritten as a closed set of first 
order differential equations for $G_1$ and $G_2$,  which turn out to be
soluble in terms of parabolic cylinder functions. Instead of working 
with $G_1$ and $G_2$,      
it is convenient to define the ratios
\bg
R_n = {1\over 2n + 1} {G_{2n+2}\over G_{2n}}
\label{f}
\ed
for $n$ even,  and
\bg
Q_n = {1\over 2n}  { G_{2n+1} \over G_{2n-1}  }
\label{g}
\ed
for $n$ odd.  
We shall set $\mu = 1$ for convenience.  
It can be reintroduced later 
via the Schwinger action principle for ${\del\over\del \mu} Z$,
or equivalently
by taking ${g \over \mu^2}$ to be dimensionless and $\phi$ to have
dimensions of ${1\over\sqrt{\mu}}$. 
In terms of the $R_n$, the Schwinger-Dyson equations read
\bg
(2n+1) g R_n R_{n-1} + R_{n-1} -1 =0
\label{Rdyson}
\ed
A combination of the Schwinger action principle and the Schwinger
Dyson equations
yields the Ricatti equation
\bg 
4g^2{\del R_n\over \del g}  + (1+2g) R_n + (2n+1) g R_n^2 -1 = 0
\label{h}
\ed
which by the change of variables, 
\bg
R_n = -{2t\over 2n + 1} {d\over dt} 
ln \left( e^{  {1 \over 4} t^2} f(t) \right) 
\label{i}
\ed
where $t= {1\over \sqrt{2g} }$,  leads to the defining equation of the 
parabolic cylinder functions;
\bg
f^{\prime\prime}(t) - ( {1\over 4} t^2 + n ) f(t) = 0
\label{j}
\ed
This equation has two independent real solutions $U(n,t)$ and
$V(n. t)$.  We give the definitions and some useful properties of
these functions in the next section.  One may also use $U(n,t)$ and
$U(n,-t)$  as an independent set.  Therefore we have
\bg
f(t) = \alpha_n U(n,t) + \beta_n U(n, -t)
\label{k}
\ed
which yields 
\bg
R_n = -t{ -\alpha_n U(n+1,t) +\beta_n U(n+1, -t)
	  \over \alpha_n U(n, t) + \beta_n U(n, -t)   }
\label{l}
\ed
The Schwinger-Dyson equations relating $R_n$ to $R_{n+1}$  constrain
the coefficients $\alpha_n$ and $\beta_n$,  so that we get
\bg
R_n = t { U(n+1,t) + (-1)^{n+1} \rho U(n+1, -t)
	  \over U(n, t) + (-1)^n \rho U(n, -t)   }
\label{rat}
\ed 
The parameter $\rho$ is not fixed and  sets one of the two 
boundary conditions
of the Schwinger-Dyson equations.

Since $G_0 = 1$,  the even Green's functions are given by 
\bg
G_{2n} = (2n - 1)!! t^n 
	{  U(n,t) + (-1)^{n} \rho U(n, -t) \over
	   U(0,t) + \rho U(n, -t)     } 
\label{m}
\ed
One can compute $Q_n$ in the same way we have computed $R_n$
giving,
\bg
Q_n = t{ U(n+{1\over 2},t)+ (-1)^{n+1} \omega {1\over n!}V(n+{1\over 2},t) 
         \over 
	 U(n-{1\over 2},t)+ (-1)^{n} 
			\omega {1 \over (n-1)! } V(n-{1\over 2},t)  }
\label{n}
\ed
However in this case $\omega$ is not a free parameter.    
The Schwinger-Dyson equation 
$gG_3 + G_1 =0$,  gives   $Q_1 = -{1\over 2g} = -t^2 $, and 
$\omega = \infty$,  so that  
\bg
Q_n = -{t \over n} {V(n+{1\over 2},t) \over V(n-{1\over 2}, t) }.
\label{o}
\ed
To get the odd Green's functions from $Q_n$  one also needs to know  $G_1$.
Due to the  Schwinger action principle,
\bg
{\del G_1\over \del g} = -{1 \over 4} G_5 + {1 \over 4} G_4 G_1
\label{p}
\ed
Using $G_5 = 8 Q_1 Q_2 G_1$  and the above solutions for the even
Green's functions,  we find that
\bg
{\del G_1 \over \del t} = 
\left(    (t + {1\over 2t}) + {1\over 2} 
          {  U(1,t) - \rho U(1, -t) \over U(0,t) + \rho U(0, -t)  }
\right)
G_1
\label{q}
\ed
Writing 
\bg
G_1 = { f(t) \over U(0,t) + \rho U(0, -t) }
\label{r}
\ed
One gets, 
\bg
{f{\prime}\over f} = {t \over 2} + {1\over 2t}
\label{s}
\ed
so that 
\bg
G_1 = \alpha {  t^{1\over 2} e^{t^2\over 4} \over
	        U(0,t) + \rho U(0,-t)  }
\label{t}
\ed
Thus the odd Green's functions are given by 
\bg
G_{2n+1} = {2n !!\over n!} (-t)^n 
		{V(n+{1 \over 2},t)\over V({1 \over 2},t)}
	   \alpha {  t^{1\over 2} e^{t^2\over 4} \over
	        U(0,t) + \rho U(0,-t)  } 
\label{u}
\ed
Since these are normalized Green's functions, the values of 
$\alpha$ and $\rho$ are related to the ratios of the 
coefficients of
the integral contours.  
It is amusing to note 
that there are classes of contours among which the even Green's functions
are the same,  and the odd Green's functions differ by an overall 
normalization.  These are obtained by varying $\alpha$  
with $\rho$ fixed.  This is a reflection of the fact that for the symmetric
action, the Schwinger
Dyson equations do not couple the even with the odd 
Green's functions.
The coupling comes only from the action principle for the 
odd Green's functions,
\bg
{\del\over\del g} G_{2n+1} - {1\over 4} G_4 G_{2n+1} + 
{1 \over 4} G_{2n+5} = 0
\label{v}
\ed
One can change the odd Green's functions by a global factor without 
effecting the above equations or the linear Schwinger-Dyson equations
for the odd Green's functions.

\subsection{Truncated series and the path integral solution}

It has already been noted by several authors (\cite{cooper}, 
\cite{caianello}) that 
solving the zero dimensional $\phi^4$ Schwinger-Dyson equations 
\bg
(g {{d^3   }\over {d {J^3}}}+ \,\mu \, {d  \over d J} - \, J) Z(J) = 0
\label{ordin}
\ed
by a truncated Taylor series leads to a solution $Z^T$ with
smooth limit $g \to 0^{+}$ for $M^2 > 0$. It is straightforward
to check that truncation causes odd coefficients in the series vanish. From
(\ref{Rdyson}), one obtains a continued fraction expansion for
the ratios $R_k$
\bg
R_{k}\,=\, {1 \over {1 + (2\,k+1)\,g\,R_{k+1}}}
\qquad k  = 1,2 \ldots , N-1  \qquad , \quad
R_{N}=0 \qquad , \quad  N \to \inf
\label{cf}
\ed
\bg
R_{n}\,=   {1 \over\displaystyle 1+
           { (2\,n+1)\,g \over \displaystyle{1+
             {(2\,n+3)\,g \over \displaystyle{1+
             {(2 \,n+5)\,g \over \displaystyle{1+ \ldots}}}}}}}
\label{expectedfrac}
\ed
A truncated series provides then a rational (Pad\'e) approximation for
the Green's functions, determined by the classical
convergents of the continued fraction (\ref{expectedfrac}).
By Pincherle's theorem, (\ref{expectedfrac}) converges
to the minimal solution of (\ref{rec}), the unique
solution $G_m \not = 0$ such that
$$
\lim_{m \to \inf} {G_m \over G^{\prime}_m} = 0
$$
for any other linearly independent solution $G^{\prime}_m$.
In fact, one can also prove
\cite{benderbook} that (\ref{expectedfrac})
converges to a Stieltjes function: a function $F(z)$
admitting the following integral representation
\bg
F(z)\,=\,\int_{0}^{\inf}\,{d t \,{\rho (t) \over 1+z t}}
\simeq \sum_{n=0}^{\inf} \sigma_n (-z)^n \qquad , \quad
\sigma_n = \int_0^{\inf} dt \,t^n \rho(t)
\ed
with $\rho(t) \ge 0 \quad \forall t \in [0, \inf] \,$. 
The series is asymptotic for $z \to 0$
and $-\pi < {\rm{arg(z)}} < \pi$ and the
coefficients $\sigma_n$ are moments of a positive density
function $\rho(t)$. Moreover, $F(z)$ is the Borel sum of the asymptotic series
if this sum exists.

It is straightforward to prove that
\bg
R_n = t { U(n+1,t) \over U(n, t) }
\label{w}
\ed
that is, equation (\ref{rat}) with $\rho = 0$. 
Note that $R_n \in [0,1]$, where the maximum occurs at $g =0 $.
Taking into account the relationship of the Green's functions with the
ratios $R_n$, and using the integral representations of the parabolic
functions given in the next section, one finds that
the truncated solution $Z^T$ converges to the usual path integral solution
\bg
Z^{T}(J)=
{
{\displaystyle{
 \int_{-\inf}^{+ \inf} d \phi \, \,e^{(
-{1 \over 2} \mu \, \phi^2 -{1 \over 4} g \, \phi^4 + J \phi)}} }
\over
{\displaystyle{
\int_{-\inf}^{+ \inf} d \phi \,e^{ (-{1 \over 2} \mu \,
\phi^2 -{1 \over 4} g \, \phi^4)} }} }
\label{interep2}
\ed
which, as stated,  is the unique solution 
of (\ref{ordin}) with smooth limit $g \to 0^+$ for $\mu > 0$.

\subsection{Parabolic Cylinder Functions}

The parabolic cylinder functions
$U(a,t)$ and $V(a,t)$ are independent
(real) solutions to the equation
\bg
{d y \over d t^2} -( {1 \over 4} t^2 +a) y =0
\label{pareq}
\ed
These function satisfy
\bg
U(a,t)= 2^{-{1 \over 4} -{a \over 2}} e^{-{1 \over 4} t^2} \,
\Psi({1 \over 2} a + {1 \over 4}, {1 \over 2}, {1 \over 2} t^2)
\label{relat0}
\ed
\bg
\pi V(a,t)= \Gamma(a +{1 \over 2})( \sin \pi a U(a,t)+U(a,-t) )
\label {relat1}
\ed
where $\Psi$ is a hyper-geometric function defined in
\cite{Abra}, pp. 503-535.
\bg
U(a,t)= {1 \over \Gamma(a+ {1 \over 2})}
{\displaystyle{e^{-{1 \over 4} t^2}} }
\int_0^{\inf} {
{\displaystyle{e^{-{ts-{1 \over 2}s^2}}}}\,
{\displaystyle{s^{a-{1 \over 2}}}} \, d s }
\qquad \qquad a > -{1 \over 2}
\label{relat2}
\ed
The change of variables $s=\phi^2/ 2 t$ transforms (\ref{relat2})
into
\bg
U(a,t)= {2 \over \Gamma(a+ {1 \over 2})}
({1 \over 2 \, t})^{ a + {1 \over 2} }
{\displaystyle{e^{-{1 \over 4} t^2}} }
\int_0^{\inf} {
{\displaystyle{e^{-{{1 \over 2} \phi^2
-{1 \over 4} \lm \phi^4}}}}\,
{\displaystyle{\phi^{2 a}}} \, d \phi }
\qquad \qquad {1 \over 2} \,t^2= {1 \over 4 \, \lm}
\label{relat21}
\ed
Properties of these functions are listed, for example,
in \cite{Abra}, pp. 685-720,
and \cite{benderbook}. Some useful relations are
\bg
U^{\prime}(a,t)+{1 \over 2} t \,U(a,t) + (a + {1 \over 2}) U(a+1,t)=0
\ed
\bg
V^{\prime}(a,t)+{1 \over 2} t \,V(a,t)  - V(a+1,t)=0
\qquad \qquad \qquad
\label{relat3}
\ed
\bg
U(a+2,t) = {  U(a,t) - t U(a+1,t) \over a+{3 \over 2}  }
\label{y}
\ed
\bg
U(a,0)={\sqrt{\pi} \over
2^{{1 \over 2} a + {1 \over 4}}
\Gamma({1 \over 2} a + {3 \over 4} )}
\label{relat31}
\ed
When $k$ is a non negative integer
\bg
\nonumber
V(k + {1 \over 2}, t ) =2 ^{-{k \over 2}} e^{{t^2 \over 4}}
H_k^{*}({t \over \sqrt{2}})
\qquad \qquad H_k^{*}(t) = e^{-t^2} { d^k \over d t^k} e^{t^2} \\\\
U(k + {1 \over 2}, t ) = {1 \over k!}
e^{{t^2 \over 4}} \int_t^{\inf} \, dz (t-z)^k \, e^{-{1/2} z^2}
\qquad \qquad \qquad  \qquad
\label{relat4}
\ed
Asymptotic expansions for large $t$
\bg
U(a,t) \simeq e^{-{1 \over 4}t^2} t^{-a - {1 \over 2}}
(1 -{(a+{1 \over 2})(a+{3 \over 2}) \over 2\, t^2}
+{(a+{1 \over 2})(a+{3 \over 2}) (a+{5 \over 2})(a+{7 \over 2})
\over 2\dot4 \, t^4} - \ldots) 
\label{z}
\ed
for $|{\rm{arg(t)}}| < {1 \over 2} \pi$
\bg
V(a,t) \simeq {\sqrt{2 \over \pi}}
e^{{1 \over 4}t^2} t^{a - {1 \over 2}}
(1 +{(a-{1 \over 2})(a-{3 \over 2}) \over 2\, t^2}
+{(a-{1 \over 2})(a-{3 \over 2}) (a-{5 \over 2})(a-{7 \over 2})
\over 2\dot4 \, t^4} + \ldots)
\label{aa}
\ed
for $|{\rm{arg(t)}}| < {1 \over 2} \pi$

\section{Appendix II}

\subsection{Borel re-summation and exotic solutions}

Here we demonstrate the relation between various Borel re-summations and 
the exotic solutions of the Schwinger-Dyson equations
for an arbitrary polynomial 
action in a zero dimensions, $S(\phi) = {g_n\over n} \phi^n$.
It is shown that the by a suitable choice of integration contour in 
the Borel variable,  one obtains an exact solution of the Schwinger-Dyson
equation which satisfies the Schwinger action principle.
The loop  expansion about the
the classical solution $\bar\phi_{\alpha}$ yields the following 
contribution to 
the generating function; 
\bg 
Z_{\alpha}\equiv \sqrt{  {\pi \hbar \over S^{\prime\prime}  
( \bar\phi_{\alpha} ) } } 
e^{-{1\over \hbar}S(\bar\phi_{\alpha})} 
c_n \hbar^n
\label{works}
\ed
We shall assume that all the coefficients $c_n$are finite.  This precludes 
actions with 
flat directions,  $S^{\prime\prime} \ne 0$.
Note that one obtains the 
same series starting from any integration contour for which $e^{-S}$
has constant phase, and which passes through $\phi = 
\bar\phi_{\alpha}$ but no other
classical solution of equal or lower action. 
This series is asymptotic,  but its Borel transform defined 
by 
\bg
B_{\alpha}(t) = \sqrt{ {\pi\over\ S^{\prime\prime} (\bar\phi_{\alpha} ) } } 
\sum_n {c_n\over \Gamma (n+{1\over 2} ) } t^n
\label{relbor}
\ed
has a finite radius of convergence.  
In the zero dimensional theory it
converges to 
\bg
B_{\alpha}(t) = 
\sqrt{t} \oint_C d\phi{1\over t- ( S(\phi)- S(\bar\phi_{\alpha}) ) } 
\label{subrl}
\ed
where in the vicinity of $t=0$ the contour $C$ encloses, 
in the opposite sense,  
the two poles $\phi_{\alpha,i} (g_k, t)$ which coalesce to 
$\bar\phi_{\alpha}(g_k)$ at $t=0$.  
All the other poles are taken to lie outside the contour.  The 
Borel transform has a singularity when one of the exterior poles 
coalesces with one
of the interior poles,  which occurs when $t$ is equal to the action of a
neighboring classical solution, 
$t=S(\bar\phi_{\alpha^{\prime}})$.
Doing the $\phi$ integral, we may also write, 
\bg
B_{\alpha}(t)= \sqrt{t}\sum_{i=1,2} (-1)^i 
{1\over S^{\prime}(\phi_{\alpha,i}(t))}
\label{als}
\ed
where $\phi_{\alpha,i} (t)$  are the two poles which coalesce at $t=0$ ;
$\phi_{\alpha,i}(0) = \bar\phi_{\alpha}$. 

Thus far everything we have said is  standard\cite{thooft}. 
We now exhibit an exact 
relation between the Borel re-summations and the exotic 
solutions of the Schwinger-Dyson
equations.
We invert the Borel re-summation 
by writing 
\bg
Z_{\alpha} = e^{-{1\over h}S(\bar\phi_{\alpha}) } \int dt 
e^{ -{t\over \hbar } }
             { B_{\alpha}(t)\over \sqrt{t} }  
\label{sqig}
\ed
with an as yet unspecified integration contour in the complex $t$ plane.
We set $h = 1$ in what follows.
If $Z_{\alpha}$ satisfies both the Schwinger-Dyson equations and the 
action principle
then it is annihilated by the operators;
\bg
\hat L = \sum_{n=2}(n-1)g_n \del_{g_{n-1}} - g_1 
\label{sdandsac}
\ed
and
\bg
\hat H_n = {\del\over\del g_n} - {1\over n}{\del^n \over\del g_1^n}
\label{sdic}
\ed
It is convenient to define the quantity 
\bg
F_{\alpha} = \int dt e^{-t} { B_{\alpha}(t)\over \sqrt{t} }
\label{fdef}
\ed
If $\hat L Z_{\alpha} = 0$  then  $ \hat{\cal L} F_{\alpha} =0$
where 
\bg 
\hat{\cal L} =  \sum_{n=2}(n-1)g_n D_{g_{n-1}} - g_1
\label{cov}
\ed
and 
\bg 
D_{g_{n-1}} \equiv
{\del\over\del g_{n-1}} - {\del\over\del g_{n-1}}S(\bar\phi)
\label{covar}
\ed
Before proceeding,  we list several simple but useful identities. 
Due to the equation of motion, $S^{\prime} (\bar\phi_{\alpha}(g_k) ) = 0$,
one has  
\bg
{\del\over\del g_n}S(\bar\phi_{\alpha}) = {1\over n}{\bar\phi_{\alpha}}^n
\label{erg}
\ed
Another obvious identity is   
\bg
\phi_{\alpha,i}(t=0, g_k)= \bar\phi_{\alpha}(g_k)
\label{alsobv}
\ed
Two other identities are found by differentiating the relation 
$t = S(\phi_i(t,g_k)) - S(\bar\phi(g_k))$  with respect to $t$ and $g_n$:
\bg
{\del\over\del t}\phi_{\alpha,i}(t, g_k)= 
{1\over S^{\prime}(\phi_{\alpha,i}(t,g_k) ) }
\label{jig}
\ed
\bg
{\del\over\del g_n}\phi_{\alpha,i}(t,g_k) = 
{  {1\over n}(\phi^n_{\alpha,i}(t,g_k) - \bar\phi^n_{\alpha}(g_k) )
    \over S^{\prime} ( \phi_{\alpha,i}(t,g_k) )  }
\label{wuc}
\ed
The equations of motion are used again in deriving the last equation.
Using these identities the quantity
\bg
\hat{\cal L} F= \int dt e^{-t}(-1)^i {\del\over \del t}
[\sum_{n=2}(n-1)g_n D_{g_{n-1}} - g_1]\phi_{\alpha,i}(t)
\label{qant}
\ed
may be rewritten, after much algebra, as 
\bg
\int dt {\del\over\del t}[  \sum_i e^{-t} (-1)^i 
\sum_n g_n {\bar\phi}_{\alpha}^{n-1}{\del\over\del t}\phi_{\alpha,i}(t)   ] 
\label{rerit}
\ed
which vanishes for {\it {any}} contour in $t$ via the equations of motion
$\sum_n g_n {\bar\phi}_{\alpha}^{n-1} =0$.  Note that it was also not 
necessary to
encircle the zeroes of $t-S(\phi)+S(\bar{\phi}_\alpha)$ in any particular 
way when doing
the initial $\phi$ integral.  The factor of $-1^i$ was certainly not 
required to satisfy
the equation $\hat L Z = 0$. Note however that this equation is some 
combination of the 
Schwinger-Dyson equation and the action principle.
We must also see if the action principle is separately satisfied.  To this end,
consider the quantity
\bg
[klD_{g_l}D_{g_k} + (k+l)D_{g_ {k+l} }]F =
\int dt e^{-t}{\del\over\del t}
\sum_i(-1)^i[klD_{g_l}D_{g_k} + (k+l)D_{g_ {k+l} }]\phi_{\alpha,i}(t)
\label{yip}
\ed
which will vanish if the action principle is satisfied.
Using the same identities this quantity may rewritten as
\bg
\int dt {\del\over\del t} \left[ e^{-t} \sum_i (-1)^i 
kl{\del\over\del g_k}{\del\over\del g_l}\phi_{\alpha,i}(t) \right] =
\left. \sum_i (-1)^i e^{-t} kl{\del\over\del g_k}
{\del\over\del g_l}\phi_{\alpha,i}(t) 
\right| _{\del\Gamma}
\label{hur}
\ed
which vanishes for several choices of contours in the $t$ plane.
The contour may be closed around some Borel singularities,  or it may
begin and end at 
$Re( t) = +\infty$ winding around some singularities at finite $t$. 
Alternatively the contour may begin at $t=0$ and end at $Re( t) = +\infty$.
The latter choice is viable because at $t=0$ the two poles $\phi_{\alpha,i}(t)$
coalesce and the factor of $\sum_i (-1)^i$ then causes the $t=0$ boundary term
to vanish.   As long as some such contour is used,  then one obtains an exact
solution of the Schwinger-Dyson 
equation satisfying the 
action principle.  
We suspect that the analysis above generalizes to 
theories with a greater number degrees of freedom, provided there are no
flat directions, though as yet we have not constructed a proof.

\end{document}